\documentclass[reprint,
superscriptaddress,
showpacs,
 amsmath,amssymb,
 aps,
 prl,
]{revtex4-2}

\usepackage[english]{babel}
\usepackage{graphicx}% Include figure files
\usepackage{dcolumn}% Align table columns on decimal point
\usepackage{bm}% bold math
\usepackage{hyperref}% add hypertext capabilities
\usepackage[mathlines]{lineno}% Enable numbering of text and display math
\usepackage{xcolor}
\usepackage{tikz}
\usetikzlibrary{patterns}
\usepackage{natbib}
\usepackage{braket}
\usepackage{bbold}

\usepackage{cleveref}
\usepackage{longtable}
\usepackage{float}
\usepackage{bm}
\usepackage{tikz}
\usetikzlibrary{calc,patterns,decorations.pathmorphing,decorations.markings}
\PassOptionsToPackage{hyphens}{url}

\newcommand{\Fop}[3]{F_{#1}^{#2}(#3)}

\newcommand{\hGamma}{\hat{\Gamma}^{(2)}}

\newcommand{\qte}[1]{\textquotedblleft{#1}\textquotedblright}

\begin{document}
\title{Symmetry oscillations in strongly interacting one-dimensional mixtures}

\author{S. Musolino}
\affiliation{Universit\'e Grenoble Alpes, CNRS, LPMMC, 38000 Grenoble, France}
\affiliation{Universit\'e C\^ote d'Azur, CNRS, Institut de Physique de Nice, 06200 Nice, France}
\author{M. Albert}
\affiliation{Universit\'e C\^ote d'Azur, CNRS, Institut de Physique de Nice, 06200 Nice, France}
\affiliation{Institut Universitaire de France}
\author{A. Minguzzi}
\affiliation{Universit\'e Grenoble Alpes, CNRS, LPMMC, 38000 Grenoble, France}
\author{P. Vignolo}
\affiliation{Universit\'e C\^ote d'Azur, CNRS, Institut de Physique de Nice, 06200 Nice, France}
\affiliation{Institut Universitaire de France}

\begin{abstract}
Multicomponent quantum mixtures in one dimension can be characterized by their symmetry under particle exchange.  For a strongly interacting Bose-Bose mixture, we   show that the time evolution of the momentum distribution  from an initially symmetry-mixed state is quasiconstant for a SU$(2)$ symmetry conserving Hamiltonian, while it displays large oscillations in time  for the symmetry-breaking case  where inter- and intraspecies interactions are different.  
Using the property that the momentum distribution operator at strong interactions commutes with the class-sum operator, the latter  acting as a symmetry witness, we show that the momentum distribution oscillations correspond to symmetry oscillations, with a mechanism analogous to  neutrino flavor oscillations.   
\end{abstract}
\maketitle

In quantum mechanics, when the initial state is not an eigenstate of the time-evolving operator, its time evolution displays oscillations, signalling  quantum coherence of the state. From Rabi oscillations among single-particle states to Josephson oscillations in an emerging effective two-level system, such oscillations are used for a wealth of applications: from qubits  \cite{devoret2004} to metrology standards \cite{rev:pezzequantum}. Quantum mechanical oscillations may occur as well among other degrees of freedom, for example, neutrino oscillations are due to the fact that the flavor basis in which neutrinos are defined does not coincide with the basis of their evolution~\cite{art:neutrinos_pdg}. In this Letter, we focus on symmetry oscillations, i.e.  oscillations among states with a defined symmetry under particle exchange. Particle exchange symmetry is a crucial property of many-body systems: while the wave function of indistinguishable bosons (spin-polarized fermions) is totally symmetric (antisymmetric) under exchange, the wave function of multicomponent mixtures has a  symmetry under exchange among different types of particles that is not {\it a priori} specified~\cite{art:sutherland-BF}.

Strongly interacting  mixtures in one dimension (1D) represent a special class of systems where symmetry aspects can be unambiguously investigated.  The versatility and controllability of ultracold atoms made it  possible to reach the strongly interacting regime in experiments both for bosonic and fermionic particles ~\cite{art:kinoshita2004,art:paredes2004,art:fermionizations-fermions-jochim}, and several aspects have been studied ever since, both at equilibrium~\cite{art:kinoshita_PRL2005, art:Haller_2009, art:fluctuations_bouchoule2011} and out of equilibrium~\cite{art:newtoncradle_kinoshita2006,  art:dyn_ferm_wilson2020, art:ghd_bouchoule2022}.  
Theoretically, 1D systems may be described by a wealth of complementary methods, both analytical, as bosonization and exact solutions~\cite{book:giamarchi_1D, rev:Sowinski2019, art:minguzzi2022strongly, rev:mistakidis}, and numerical~\cite{rev:DMRG2005, rev:MPS_Cirac2021}. Strongly repulsive contact interactions  give rise to a gas of impenetrable particles whose wave function vanishes when two particles approach each other. This corresponds to the celebrated  Tonks-Girardeau (TG)~\cite{art:tonks1936,art:Girardeau1960} regime, first discovered for single-component bosons and then extended to multicomponent mixtures~\cite{art:girardeau_minguzzi2007, art:deuretzbacher2008_lett,art:volosniev_nat}.  In the TG limit, the many-body wave function of a mixture factorizes in an orbital and spin part. The orbital part can be exactly  written by mapping onto the one of noninteracting fermions in the same external potential, while, for a two-component mixture,  the spin part is described by the Hamiltonian of an effective  spin chain, where the site index corresponds to particle index and hopping coefficients are fixed by the gradients of the many-body wave function upon particle exchange~\cite{art:deuretzbacher2008_lett, art:volosniev_nat}. A similar mapping has also been extended to SU$(N)$ mixtures~\cite{art:decamp_sym2016,art:decamp2016_high}. Because of particle-exchange symmetry, a given many-body state of a mixture  can be classified in terms of   irreducible representations of the symmetry group of permutations of $N$ elements, $S_N$, each associated to a different Young diagram~\cite{book:hamermesh, art:fang2011, art:Harshman_2015, art:decamp2016_high, art:aupetit}.
  Suitable symmetry witness operators are the class-sum operators~\cite{art:Katriel_1993}, which probe particle permutation cycles. 
  While spin degrees of freedom do not affect the total density of the system,
   which coincides with the one of the mapped Fermi gas,  the spatially nonlocal correlation functions, such as the one-body density matrix  and, consequently, the momentum distribution of the interacting mixture differ considerably from those of a noninteracting Fermi gas.

In this Letter, we focus on the momentum distribution of a strongly interacting 1D Bose-Bose mixture, and show that it is an ideal observable to probe the symmetry of the state, since the momentum distribution  operator and the symmetry witness can be simultaneously diagonalized independently of the number of components.   We monitor both its peak at zero momentum, which characterizes the presence of quasi-off-diagonal long-range order,  and  its tails at large momenta, which show a universal power law decay and give information on short-range correlations. We propose a quench protocol in which, starting from an initial  state with several symmetry components, we let the state evolve and follow the dynamics of the  momentum distribution. 
Focusing specifically on the two-component case, we show that the time evolution is very different depending on whether the system Hamiltonian is SU$(2)$ symmetric, i.e. with equal inter- and intraspecies interactions, or symmetry breaking (SB), i.e., with different inter- and intraspecies interactions. The momentum distribution remains approximately constant in time for the SU$(2)$ case, while it  oscillates for the symmetry-broken case. In both cases, the dynamical evolution of the momentum distribution is very close to the one of the symmetry witness.

\textit{Model.}-- 
We consider a 1D mixture of $N$ bosonic particles with two components ($\uparrow$ and $\downarrow$) with $N= N_\uparrow+ N_\downarrow$, of   mass $m$ interacting via repulsive contact interactions as described by the Hamiltonian
\begin{equation}
\begin{split}
H&=\sum_{\sigma=\uparrow,\downarrow} \sum_{i=1}^{N_\sigma} 
\left[-\frac{\hbar^2}{2m}\frac{\partial^2}{\partial x_{i,\sigma}^2} + V(x_{i,\sigma},t)\right] \\ &+g_{\uparrow\downarrow}\sum_{i, j}\delta(x_{i,\uparrow}-x_{j,\downarrow})  
+ \sum_{\sigma= \uparrow, \downarrow} g_{\sigma\sigma}\sum_{i<j} \delta(x_{i,\sigma}-x_{j,\sigma}),
\end{split}
 \label{eq:ham}
\end{equation}
 where $V(x,t)$ is a spin-independent external trapping potential, and $g_{\sigma\sigma'}$ is the inter- ($\sigma\neq\sigma'$) or intraspecies ($\sigma=\sigma'$) interaction strength.  In the limit of $g_{\sigma\sigma'}\rightarrow +\infty$, for any $\sigma,\sigma'$, the  many-body wave-function $\Psi$ vanishes whenever $x_{i,\sigma}=x_{j,\sigma'}$. Using a generalized time dependent Bose-Fermi mapping~\cite{art:girardeau-timedependent,art:deuretzbacher2008_lett, art:volosniev_nat}, the time-evolving exact many-body wave function can be  written  as
  \begin{equation}
  \Psi(\vec{X}, \vec{\sigma}, t)= \sum_{P\in S_N}   \braket{\vec{\sigma} |\hat{P}|\chi(t)} \theta_P(\vec{X})\Psi_A(\vec{X},t),
  \label{eq:Psi_MB}
  \end{equation}
which allows us to describe a system with spatial ($\vec{X} \equiv \{ x_{1},\dots,x_{N} \}$) and spin ($\vec{\sigma} \equiv \{ \sigma_{1},\dots,\sigma_{N} \}$) degrees of freedom where we have set $\{x_i,\sigma_i\}=x_{i,\sigma}$ if $\sigma=\uparrow$ and  
$\{x_{i+N_\uparrow},\sigma_{i+N_\uparrow}\}=x_{i,\sigma}$ if $\sigma=\downarrow$ the spatial coordinate and the spin value of the $i$th particle. The operator $\hat P$ in Eq.~\eqref{eq:Psi_MB} corresponds to a permutation in $S_N$, $\ket{\chi(t)}$ is the spin state at time $t$, $\theta_P(\vec{X})$ is the generalized Heaviside function, which is
equal to 1 in the coordinate sector $x_{P(1)}<\dots<x_{P(N)}$ and 0 elsewhere, and $\Psi_A=A \Psi_F$ with $A =\prod_{i<j} \mathrm{sgn}(x_i-x_j)$. Finally, $\Psi_F=(1/\sqrt{N!}) \det [\phi_j(x_k,t)]$ is  the wave-function of $N$ spinless noninteracting fermions, built from the single particle orbitals $\phi_j(x,t)$ for a quantum particle in the potential $V(x,t)$, time evolved from an initial Fermi sphere configuration~\cite{art:girardeau-timedependent}.

In the limit of strongly repulsive interactions, Eq.~\eqref{eq:ham} can be mapped to an effective spin Hamiltonian, in a Hilbert space of dimension $M=N!/(N_\uparrow!N_\downarrow!)$, where the site index corresponds to the particle index~\cite{art:massignan2015, art:Yang_2016, art:barfknecht_2018, art:aupetit}, 
and the number of sites corresponds to the total number of particles, which for $g_{\uparrow\uparrow}=g_{\downarrow\downarrow}= g$ reads
 \begin{equation}
 \begin{split}
\hat{H}_\mathrm{s} &= -\sum_{i=1}^{N-1}  2\alpha_i \left[\frac{1}{g_{\uparrow\downarrow}} (S_{i}^{(x)}S _{i+1}^{(x)} +S_{i}^{(y)} S_{i+1}^{(y)} )\right. 
\\ &+ \left. \left(\frac{2}{g} - \frac{1}{g_{\uparrow\downarrow}}\right)S_{i}^{(z)} S_{i+1}^{(z)}+\left(\frac{1}{4 g_{\uparrow \downarrow}} + \frac{1}{2 g} \right) \mathbf{I}\right], 
 \label{eq:Hspin}
 \end{split}
 \end{equation}
 where $\mathbf{I}$ is the identity matrix, $\mathbf{S}_i = (S_i^{(x)}, S_{i}^{(y)}, S_{i}^{(z)})$ is the spin matrix acting at site $i$ in terms of its projections, and the couplings $\alpha_i$ are given by
 \begin{equation}
 \alpha_i = \frac{ N! \hbar^4}{m^2} \int d\vec{X} \theta_\mathrm{id}(\vec{X})\delta(x_i - x_{i+1}) \Big|\frac{\partial \Psi_A}{\partial x_{i}}\Big|^2. 
 \label{eq:Ji}
 \end{equation}
 Notice that  if $g=g_{\uparrow\downarrow}$ the effective spin Hamiltonian (\ref{eq:Hspin}) is a SU$(2)$-symmetric $XXX$ Heisenberg chain $\hat{H}_\mathrm{SU}$, while in the general case $g\neq g_{\uparrow\downarrow}$ it corresponds to a $XXZ$  chain $\hat{H}_\mathrm{SB}$, which breaks the SU$(2)$ symmetry. 
 The time evolution of the spin part of the many-body wave function is given by  $\ket{\chi(t)} = e^{-i \hat{H}_s t} \ket{\chi(0)}$, where $ \ket{\chi(0)}$ is the spin configuration of the initial state. We remark that the mapping can be extended to multicomponent systems with $\kappa > 2$  components~\cite{art:deuretzbacher2014} by employing the corresponding generators of the SU$(\kappa)$ group~\cite{art:sutherland1975}.

\textit{Observables and symmetry witness.}-- We demonstrate here that the total momentum distribution, given by the Fourier transform of the one-body density matrix
$\rho^{(1)}(x, x', t) =  \bra{\Psi} \sum_\sigma\sum_{i=1}^N (\ket{x, \sigma} \bra{x', \sigma})_i \ket{\Psi}$ in the case of a strongly interacting mixture gives direct access to the symmetry content of the state. Indeed,  using Eq.~(\ref{eq:Psi_MB}), the total momentum distribution reads 
$n(k, t)=  \bra{\chi(t)} \hat n_k  \ket{\chi(t)}$, where the momentum density operator in spin space is given by~\cite{art:deuretzbacher2016_num}
\begin{equation}
\hat n_k = \sum_{i, j=1}^N  \hat{P}_{(i,..,j)}   R^{(i, j)}(k), 
\label{eq:nk}
\end{equation}
where $\hat{P}_{(i,...,j)}$ is the cyclic (anticyclic)  permutation $i \to i+1 \to \cdots \to j-1 \to j \to i$ ($i \to i-1 \to \cdots \to j+1 \to j \to i$) if $i<j$ ($i>j$) and the identity if $i=j$. The orbital part is given by
\begin{equation}
\begin{split}
R^{(i, j)}(k) &= \frac{N!}{2\pi}\int dx dx'  e^{-ik(x-x')} \displaystyle\int_{I_{ij}} \left(\prod_{n\neq i} dx_n \right)\\
& \Psi_\mathrm{A}(x_1, \dots, x_{i-1}, x, x_{i+1}, \cdots, x_N)\\ &\times  \Psi_\mathrm{A}(x_1, \dots, x_{i-1}, x', x_{i+1}, \cdots, x_N),
\end{split}
\label{eq:rhoij}
\end{equation} 
with $I_{ij}$ is the integration interval defined by $x_1 < \cdots < x_{i-1} < x < x_{i+1} < \cdots < x_{j-1} < x' < x_{j} < \cdots < x_N$ ($i<j$)~\cite{art:deuretzbacher2016_num}.
Notice that $R^{(i, j)}(k)$ fixes the type of permutation cycles  contributing to the momentum distribution at a given $k$.  At large $k$, only binary permutations $i\rightarrow i+1$ have nonzero weight: the only particle exchanges that matter at short distances are those involving two particles, which are responsible for building the large-momentum tails  $n(k)=\mathcal{C} k^{-4}$ where $\mathcal{C}=\mathcal{C}_T+\mathcal{C}_b$ is given by the two-body Tan contact $\mathcal{C}_T$ and,   in a box trap, additional boundary terms $\mathcal{C}_b$ \cite{art:schehr_wigner2021,art:aupetit2023}. At $k=0$, instead, all permutation cycles play a role: in order to establish quasi-off-diagonal long-range order one needs to sample the coherence established among all the particles~\cite{art:aupetit}.  We also recall that  $R^{(i, j)}(k)$ may be time dependent if the dynamical evolution involves variations of the external trapping potential.

\begin{figure}
    \centering
    \includegraphics[width=0.4\textwidth]{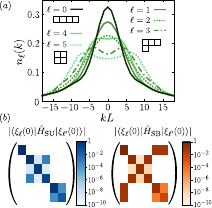}
    \caption{(a) Symmetry-resolved  momentum distribution  $n_\ell(k)$  (see Eq.~\eqref{eq:nk_decomp}) for $N=4$ bosons. The corresponding Young diagrams  are also represented in the legend. (b) Absolute value of the matrix elements of the SU$(2)$ (left) and the symmetry-breaking (right) Hamiltonian (dimensionless, normalized to the value of the largest element)  in the basis of  $|\xi_\ell(k)\rangle $ at $k=0$ in logarithmic scale. In both panels, the indices $\ell$ are sorted in a way that $\ell=0$ corresponds to the $\hat{\Gamma}^{(2)}-$eigenvalue $\gamma_0 = 6$ (the most symmetric one), $\ell=1, 2, 3$ to  $\gamma_{1, 2, 3} = 2$, and $\ell=4, 5$ to  $\gamma_{4, 5} = 0$ (the most antisymmetric). The interaction strengths are  $g=g_{\uparrow\downarrow}=20 (\hbar^2/mL)$ for the SU$(2)$ case and $1/g=0$ and $g_{\uparrow\downarrow}=20(\hbar^2/mL)$ for the SB one.  }
    \label{fig:matrices}
\end{figure}

In order to characterize the symmetry of the mixture,  
we use a set of generators of the SU$(2)$ algebra  related to 
the permutation symmetry group. In particular, we take as symmetry witness the two-cycle class sum operator, which corresponds to the group partition 
related to the sum of the transpositions $\hat{P}_{i,j}$~\cite{art:fang2011, art:decamp2016_high, art:aupetit}, namely
\begin{equation}
\hat{\Gamma}^{(2)} = \sum_{i<j} \hat{P}_{i, j}.
\label{eq:Gamma2}
\end{equation} 
The crucial observation for the following derivation is that the momentum density operator $\hat n_k$ 
has the property~\cite{suppmat}
\begin{equation}
[\hat{\Gamma}^{(2)},\hat n _k ]  = 0,
    \label{eq:comm_HJ}
\end{equation}
which is valid for an arbitrary number of components in the mixture.
Therefore, one can make the following symmetry decomposition of the momentum distribution:
\begin{equation}
n(k, t) = \sum_{\ell=1}^{M} |\braket{\chi(t)| \xi_\ell(k)}|^2 n_\ell(k),
\label{eq:nk_decomp}
\end{equation}
 where the basis $\{\ket{\xi_\ell(k)}\}$ diagonalizes both $\hat n_k$ and $\hat \Gamma^{(2)}$, such that $\hat n_k  \ket{\xi_\ell(k)} = n_\ell(k) \ket{\xi_\ell(k)}$ and $\hat{\Gamma}^{(2)} \ket{\xi_\ell(k)} = \gamma_\ell \ket{\xi_\ell(k)}$.
Figure~\ref{fig:matrices}(a) shows the symmetry-resolved momentum distributions $n_\ell(k)$ for a two-component bosonic mixture with $N_\uparrow=2$ and $N_\downarrow=2$. Interestingly, the most symmetric configuration has a single peak at $k=0$, while the most antisymmetric one -- compatible with bosonic exchange symmetry -- has as many peaks as the number of particles in each component, as it has been previously noted for fermionic mixtures~\cite{art:deuretzbacher2016_num,art:decamp_strong_2017}.
In the time evolution under a symmetry-breaking Hamiltonian, the quantum state explores different symmetry components and  the momentum distribution oscillates among the various shapes shown in Fig.~\ref{fig:matrices}(a). 

The result (\ref{eq:comm_HJ}) is nontrivial. Notice, for example, that the SU$(2)$ Hamiltonian, also  constructed with permutation operators as the symmetry witness, does not commute with $\hat n_k$, and, therefore, one cannot find a basis  that simultaneously diagonalizes $\hat{H}_\mathrm{SU}$, $\hat{\Gamma}^{(2)}$ and  $\hat n_k$~\cite{suppmat}. 
Figure~\ref{fig:matrices}(b) shows the matrix elements $\epsilon^{\ell, \ell'}(k) = |\braket{\xi_\ell(k) |\hat{H}|\xi_{\ell'}(k)}|$ for $k=0$ and $N=4$ of both the SU$(2)$ Hamiltonian and the SB Hamiltonian. 
We notice that, for  $\hat{H}_\mathrm{SU}$, the nonzero off-diagonal terms are smaller with respect to the diagonal terms and are nonzero only inside the same symmetry sector, following from the fact that $[\hat{H}_\mathrm{SU}, \hat{\Gamma}^{(2)}] = 0$. Contrarily, for  $\hat{H}_\mathrm{SB}$, not only the off-diagonal terms are comparable in magnitude to the diagonal terms, but there is also coupling between different symmetries. As a consequence, the dynamics of the momentum distribution predicted in Eq.(\ref{eq:nk_decomp}) will be very different in the SU$(2)$-symmetric and symmetry broken case.

\textit{Momentum distribution dynamics and symmetry oscillations.}-- 
In order to observe symmetry oscillations,  we solve exactly the dynamics of a strongly interacting equally balanced Bose-Bose mixture with $N_\uparrow=N_\downarrow$ in a box trap, prepared initially in a symmetry-mixed configuration, and evolving either under the SU$(2)$ Hamiltonian, i.e.  where we take $g=g_{\uparrow \downarrow}$  or the SB Hamiltonian, where we take $g \to \infty$ and  $g_{\uparrow \downarrow}$ large but finite, as in Ref.~\cite{art:aupetit}.
We choose as the initial state  
the configuration with all spin-up particles on the left and spin-down on the right, namely $\ket{\chi(0)} = \ket{\uparrow \uparrow \uparrow \downarrow\downarrow\downarrow}$, which is of experimental relevance~\cite{art:zwierlein-roati2011}.
 Since there is no change in the external potential, the dynamics occurs only among spin components,  the only time-dependent parts of Eq.~\eqref{eq:Psi_MB} are the spin coefficients, 
$\braket{\vec{\sigma} |\hat{P}|\chi(t)}$.
In real space, the dynamics gives rise to spin mixing and oscillations of the magnetization, analogous to what was observed in Ref.~\cite{art:pecci2022}  for fermionic mixtures.

\begin{figure}
    \centering
       \includegraphics[scale=0.42]{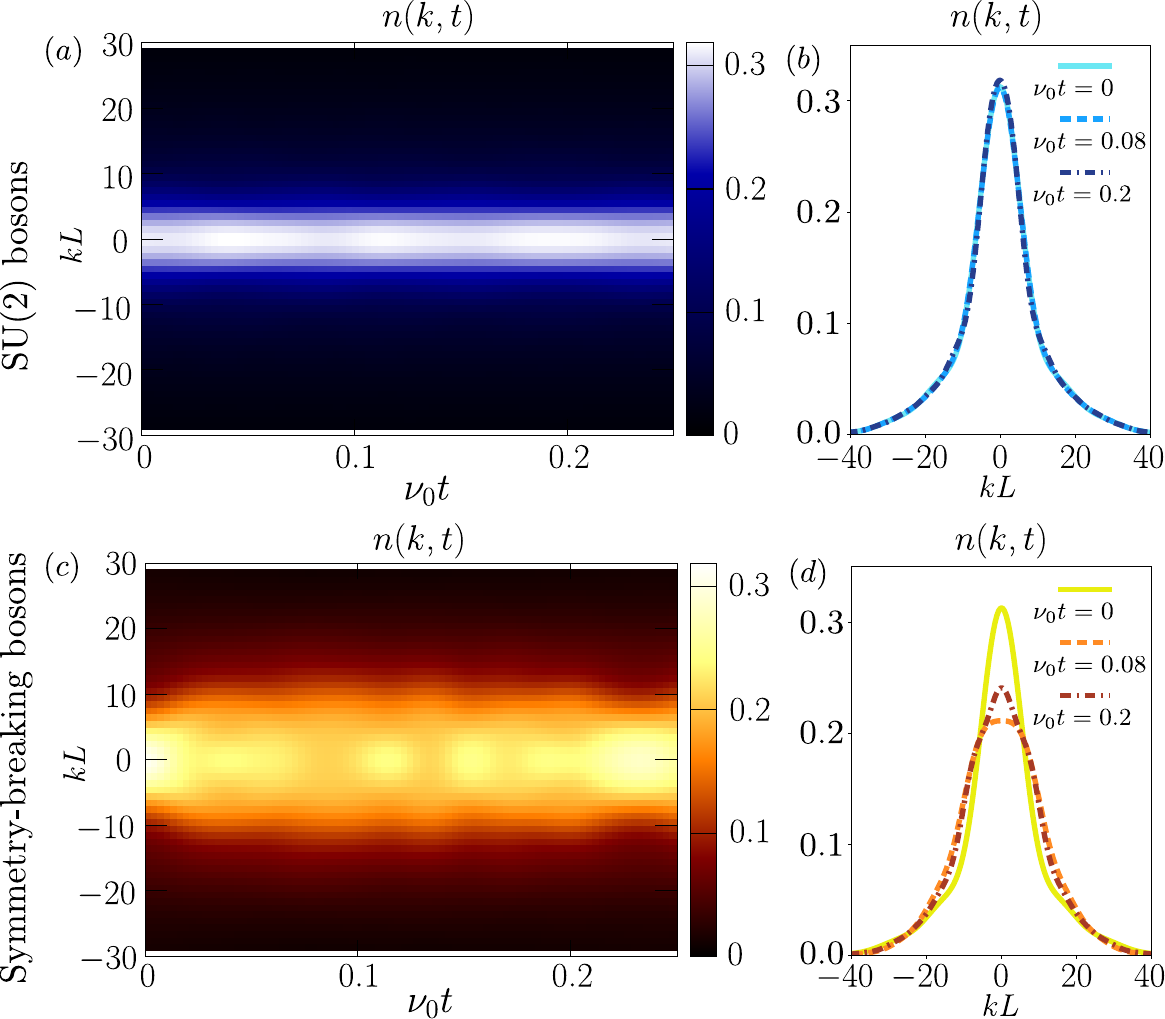}
    \caption{Time evolution of the momentum distribution $n(k, t)$ for $3+3$ SU$(2)$ bosons [(a) and (b)] and  symmetry breaking bosons [(c) and (d)]  trapped in a box with size $L$ for an initial state $|\chi(0)\rangle=|\uparrow\uparrow \uparrow \downarrow  \downarrow  \downarrow \rangle$. (a)-(c) Momentum-time map of $n(k, t)$. (b)-(d)  $n(k, t)$ at given time in the dynamics as indicated in the legend  of the figures (b) and (d)). The time is in units of $\nu_0 = \hbar/(m L^2)$ and the interaction strengths are  $g=g_{\uparrow\downarrow}=20 (\hbar^2/mL)$ for the SU$(2)$ case and $1/g=0$ and $g_{\uparrow\downarrow}=20(\hbar^2/mL)$ for the SB one. }
    \label{fig:nk}
\end{figure}

The dynamics in momentum space is illustrated in Fig.~\ref{fig:nk}, where we show the evolution in momentum-time maps and at selected times.  While in the case of the SU$(2)$-symmetric mixture the dynamics is practically constant (see below for more details), clear oscillations are seen in the momentum distribution of the symmetry-broken case. Such oscillations are particularly visible at small momenta, but actually affect the whole momentum distribution down to its tails.

\begin{figure}
    \centering
    \includegraphics[scale=0.55]{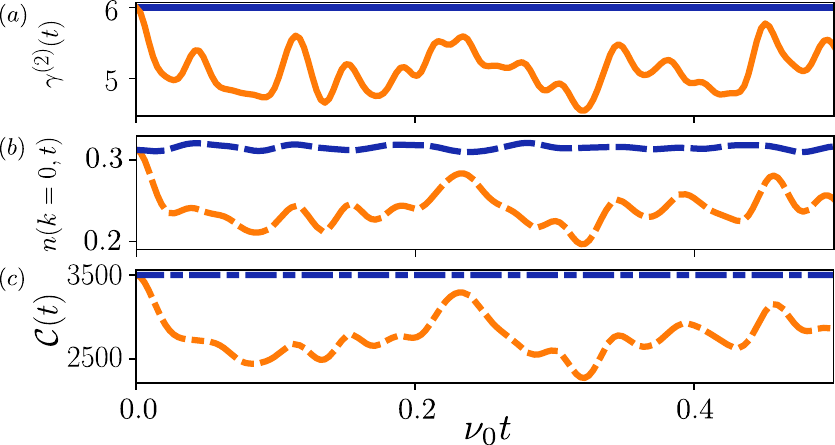}
    \caption{Time evolution of the (a)  expectation value of  $\hat{\Gamma}^{(2)}$ on the out-of-equilibrium state (Eq.~\eqref{eq:Gamma2}, solid lines), (b) momentum distribution at $k=0$ (dashed lines), and (c) weight of the large-momentum tails (dot-dashed lines). In all panels, the curves for SU$(2)$ bosons are in blue (upper curves) and for symmetry-breaking bosons in orange (lower curves).}
    \label{fig:final}
\end{figure}

To infer the relation of such oscillations with the symmetry oscillations, we follow comparatively  in Fig.~\ref{fig:final} the time evolution of the expectation value of the symmetry witness 
\begin{equation}
\gamma^{(2)}(t)= \langle \chi (t) |\hat \Gamma^{(2)}|  \chi (t) \rangle,
\end{equation}
together with the value of the momentum distribution at $k=0$ and of the weight of the large-momentum tails $\mathcal{C}(t) = \lim_{k \to \infty} k^4 n(k, t)$.  
For the reference case of a SU$(2)$-symmetric Hamiltonian, the symmetry witness is constant in time since $[\hat{H}_\mathrm{SU},\hat \Gamma^{(2)}]=0$. Also 
the Tan contact 
is constant in time, as follows from the fact that   the only nonzero term of Eq.~(\ref{eq:nk}) at large momenta is  $R^{(i, i+1)}(k)$, which implies that the Tan contact is proportional to the SU$(2)$ Hamiltonian, as readily follows by recalling that the latter can be rewritten -- taken apart a constant term which is irrelevant in our discussion -- as $\hat{H}_\mathrm{SU}=-\sum_i J_i \hat{P}_{i,i+1}$, with $J_i=\alpha_i/g$.  The peak of the momentum distribution instead displays small oscillations. These are due to the fact that $[\hat{H}_\mathrm{SU},\hat n_k]\neq 0$, and correspond to oscillations {\em within} a symmetry sector of the Hamiltonian represented in Fig.~\ref{fig:matrices}.

The dynamics of the symmetry-broken case shows a striking similarity among the time evolution of $\gamma^{(2)}(t)$, $\mathcal{C}(t)$ and $n(k=0,t)$.  
The minor differences among them is due to the fact that the symmetry weights in  
$\gamma^{(2)}(t)= \sum_{\ell=1}^{M}  |\braket{\chi(t)| \xi_\ell(k)}|^2 \gamma_\ell$ and in 
$n(k, t) = \sum_{\ell=1}^{M} |\braket{\chi(t)| \xi_\ell(k)}|^2 n_\ell(k)$
are different and in particular, the $\gamma_\ell$ may have degenerate values. The close similarity of the three curves confirms that the momentum distribution oscillations correspond to particle-exchange symmetry oscillations.

\textit{Conclusions.} -- In this work we have demonstrated that the dynamics of the total momentum distribution of a strongly interacting mixture with broken SU$(2)$ symmetry can be used to probe particle-exchange symmetry oscillations, induced by the time evolution of an initially symmetry-mixed state. 
We have chosen this initial state because it is relevant for experiments with ultracold atoms~\cite{art:wei_science2022}, but the same qualitative results are obtained for initial states with a well-defined symmetry, as it would happen in neutrino-physics experiments~\cite{suppmat}.
We have shown that the oscillations are visible both in  the zero-momentum peak and in the tails of the momentum distribution, and faithfully reflect the oscillations of the symmetry witness operator $\hat{\Gamma}^{(2)}$.  Our predictions are relevant for experiments with multicomponent ultracold atoms~\cite{art:Spielman_2014, art:Cominotti_2024}, where the interatomic interactions can be tuned using Feshbach resonances and the time evolution of the momentum distribution is accessible using time-of-flight experiments. Our result paves the way for the investigation 
of the interplay of symmetry and interactions in multicomponent quantum systems and could be extended to systems where the symmetry is broken by the presence of different masses~\cite{Harshman2017}, and different inter- and intraspecies interactions, which was previously addressed in the weakly to intermediate interacting regime~\cite{art:fogarty}.

\begin{acknowledgments}
We thank G. Aupetit-Diallo, J. Beugnon, F. Cataliotti, and J. Dalibard  for valuable discussions. We acknowledge financial support from the ANR-21-CE47-0009 Quantum-SOPHA project and the support of the Institut Henri Poincar\'e (UAR 839 CNRS-Sorbonne Universit\'e), and LabEx CARMIN (ANR-10-LABX-59-01).
\end{acknowledgments}

\bibliographystyle{apsrev4-2}
\bibliography{biblio}

\appendix

\onecolumngrid
\section{End Matter}
\twocolumngrid

\textit{Proof of Eq.~\eqref{eq:comm_HJ}.}-- In this appendix, we summarize the proof of Eq.~\eqref{eq:comm_HJ} and we refer to Ref.~\cite{suppmat} for additional details of the derivation. 
We start by re-writing the momentum distribution operator (Eq.~\eqref{eq:nk}) as 
\begin{equation}
\hat{n}_k = N \hat{1} R^{(i,i)}(k) + \sum_{m=1}^{N-1} \hat{\Gamma}_\mathrm{red}^{(m+1)},
\label{eq:nk_rewr}
\end{equation}
where we have introduced a reduced form of the class-sum operators given by
\begin{equation}
\begin{split}
\hat{\Gamma}^{(m+1)}_\mathrm{red} (k) &=  \sum_{i=1}^{N-m} (\hat{P}_{(i, i+1, \dots, i+m)} \\&+  \hat{P}_{(i+m, i+m-1, \dots, i)})R^{(i,i+m)}(k),
\end{split}
\label{eq:Gamma_red}
\end{equation} 
in terms of the cycle $i \to i+1 \to \cdots i+m \to i$ and the anticycle $i+m \to i+m-1 \to \cdots i \to i+m$ and where $R^{(i, i)}$ is the same for every $i$ and $R^{(i,i+m)}(k) =  R^{(i+m, i)}(k)$ due to parity.
The (anti)cycles can be written in terms of the swap operators $\Fop{\mu}{\nu}{i}$~\cite{book:auerbach, art:auerbach1988, art:zhang_wang} which swap a particle with flavor $\mu$ with one with flavor $\nu$ at site $i$ and satisfy the commutation relation
\begin{equation}
    \left[\Fop{\mu}{\nu}{i}, \Fop{\alpha}{\beta}{j}\right]=\delta_{i,j}\left(\Fop{\mu}{\beta}{i} \delta_{\alpha\nu}-\Fop{\alpha}{\nu}{i} \delta_{\mu\beta}\right).
    \label{eq:commF}
\end{equation}
For example, for the cycle, we can write
\begin{equation}
\begin{split}
\hat{P}_{(i, i+1, \dots, i+m)} &= \sum_{\mathrm{flav.}} \Fop{\alpha_0}{\alpha_1}{i}\Fop{\alpha_1}{\alpha_2}{i+1} \cdots \\&\Fop{\alpha_{m-1}}{\alpha_m}{i+m-1}\Fop{\alpha_m}{\alpha_0}{i+m},
\end{split}
    \label{eq:PfuncF}
\end{equation}
where the sum over \qte{flav.} means that all the flavors (labeled by the Greek letters)  are summed. 

\begin{figure}
    \centering
    \includegraphics[scale=0.8]{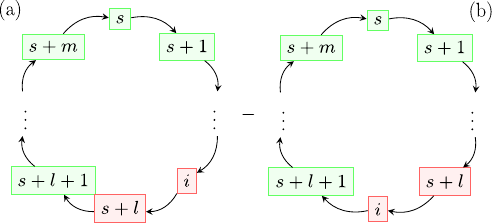}
    \caption{Schematic representation of the (a) and (b) cycles presented in the commutator $[\hGamma, \hat{\Gamma}^{(m+1)}_\mathrm{red} (k)]$ [Eq.~\eqref{eq:comm_G2}], and labeled according to the terms of Eq.~\eqref{eq:comm_proof_2}.  Every arrow represents a transposition that connects two neighboring sites, represented by the rectangles.}
\label{fig:cycleG2}
\end{figure}

While the first term of Eq.~\eqref{eq:nk_rewr}  commutes straightforwardly with $\hGamma$, the second term can be decomposed as   
\begin{equation}
\begin{split}
[\hGamma, \hat{\Gamma}^{(m+1)}_\mathrm{red} (k)] &= \sum_{i=1}^{N-1} \sum_{j=2}^{N} \sum_{s=1}^{N-m} \sum_{\mathrm{flav.}} \sum_{l=0}^m \hat{P}_{(s, \dots, s+l-1)} \\&[\Fop{\mu}{\nu}{i} \Fop{\nu}{\mu}{j},\Fop{\alpha_{l}}{\alpha_{l+1}}{s+l}] \\& \hat{P}_{(s+l+1, \dots, s+m)} R^{(s, s+m)}(k),
\end{split}
\label{eq:comm_G2}
\end{equation}
where we have used Eq.~\eqref{eq:PfuncF} for rewriting both $\hGamma$ and $\hat{\Gamma}^{(m+1)}_\mathrm{red} (k)$. The commutator in the middle of Eq.~\eqref{eq:comm_G2} can be expanded as follows
\begin{equation}
\begin{split}
      &[\Fop{\mu}{\nu}{i} \Fop{\nu}{\mu}{j},  \Fop{\alpha_{l}}{\alpha_{l+1}}{s+l}]\\&=\delta_{j, s+l} \Big(\underbrace{\Fop{\alpha_{l}}{\nu}{i} \Fop{\nu}{\alpha_{l+1}}{s+l}\delta_{\mu, \alpha_l}}_{\mathrm{(a)}} \\
        &-\underbrace{\Fop{\alpha_l}{\mu}{s+l} \Fop{\mu}{\alpha_{l+1}}{i}\delta_{\nu, \alpha_{l+1}}}_{\mathrm{(b)}}\Big)+ [i \leftrightarrow j]
        \end{split},
    \label{eq:comm_proof_2}
\end{equation} 
where $[i \leftrightarrow j]$ indicates that there are two extra terms that are the same  as the first two  with the change of variable  $i \leftrightarrow j$. By inserting Eq.~\eqref{eq:comm_proof_2} in Eq.~\eqref{eq:comm_G2}, we can see that at fixed $l$,  we obtain a difference between two cycles labeled as (a) and (b) according to Eq.~\eqref{eq:comm_proof_2} and  schematically represented in Fig.~\ref{fig:cycleG2}.

 We now notice that the (a) $l$th cycle  is equal in modulus and opposite in sign to the (b) $(l+1)$th cycle, and the same happens for  the (b) $l$th and the (a) $(l-1)$th cycles. Therefore, all the cycles with fixed $l$ cancel each other in neighboring pairs except for two. The latter, which are the (a) $l=0$ and (b) $l=m$ cycles, cancel each other because of the cycle property, namely, $\hat{P}_{(i, s, s+1, \dots, s+m)} = \hat{P}_{(s, s+1, \dots, s+m, i)}$. The same steps can be repeated for the anticycle in Eq.~\eqref{eq:Gamma_red} and for every $m$ of Eq.~\eqref{eq:nk_rewr}. Therefore, we conclude that $\hGamma$ commutes with $\hat n_k$.

\end{document}

\title{Supplemental material: Symmetry Oscillations in Strongly Interacting One-Dimensional Mixtures}

\author{S. Musolino}
\affiliation{Universit\'e Grenoble Alpes, CNRS, LPMMC, 38000 Grenoble, France}
\affiliation{Universit\'e C\^ote d'Azur, CNRS, Institut de Physique de Nice, 06200 Nice, France}
\author{M. Albert}
\affiliation{Universit\'e C\^ote d'Azur, CNRS, Institut de Physique de Nice, 06200 Nice, France}
\affiliation{Institut Universitaire de France}
\author{A. Minguzzi}
\affiliation{Universit\'e Grenoble Alpes, CNRS, LPMMC, 38000 Grenoble, France}
\author{P. Vignolo}
\affiliation{Universit\'e C\^ote d'Azur, CNRS, Institut de Physique de Nice, 06200 Nice, France}
\affiliation{Institut Universitaire de France}

\maketitle
\tableofcontents

\section{Symmetry decomposition of the momentum distribution} 
\label{sec:nk_symm}
In this Section, we show that the momentum distribution operator, $\hat{n}_k$ (Eq.~(5) of the main text), commutes with the two-cycle class-sum operator, $\hat{\Gamma}^{(2)}$ (Eq.~(7) of the main text), but not with the SU($2$) Hamiltonian, $\hat{H}_\mathrm{SU}$. 

 First, we recall the definition of $\hat{n}_k$, namely
 \begin{equation}
 \hat{n}_k \equiv \sum_{i=1}^N \sum_{j=1}^N \hat{P}_{(i, \dots, j)} R^{(i, j)}(k),
     \label{eq:nk_op}
 \end{equation}
where $\hat{P}_{(i, \dots, j)}$ is  the loop permutation defined as~\cite{art:deuretzbacher2016_num}
\begin{equation}
\hat{P}_{(i, \dots, j)} =
\begin{cases}
\hat{P}_{(i, i+1, \dots, j-1, j)} & \mathrm{if}\,\, i<j \\
\hat{1}& \mathrm{if}\,\, i=j \\
\hat{P}_{(i, i-1, \dots, j+1, j)}& \mathrm{if}\,\, i>j 
\end{cases},
\label{eq:Pi..j}
\end{equation}
which is a cyclic (anti-cyclic) permutation if $i<j$ ($i>j$) and the identity operator if $i=j$, 
 and $R^{(i, j)} (k)$ is the spatial part  defined in Eq.(6) of the main text. For convenience, we rewrite the sums over the loop permutations in Eq.~\eqref{eq:nk_op} as 
\begin{equation}
  \sum_{i=1}^N    \sum_{j=1}^N  \hat{P}_{(i, \dots, j)}  = N \times \hat{1} + \sum_{m=1}^{N-1} \sum_{i=1}^{N-m} \left(\hat{P}_{(i, i+1, \dots, i+m)} +  \hat{P}_{(i+m, i+m-1, \dots, i)} \right),
  \label{eq:Pkkpm}
\end{equation} 
where we have distinguished the diagonal part with $m=0$ ($i=j$), the cycle $i \to i+m$ ($i< j$) and its inverse $i+m \to i$ ($i> j$).
From Eq.~\eqref{eq:Pkkpm}, we now define a reduced form of the class-sum operators given by
\begin{equation}
\begin{split}
\hat{\Gamma}^{(m+1)}_\mathrm{red} (k) &=  \sum_{i=1}^{N-m} (\hat{P}_{(i, i+1, \dots, i+m)} +  \hat{P}_{(i+m, i+m-1, \dots, i)})R^{(i,i+m)}(k),
\end{split}
\label{eq:Gamma_red}
\end{equation} 
 where $R^{(i,i+m)}(k) =  R^{(i+m, i)}(k)$ due to parity. The momentum operator is therefore defined as $\hat{n}_k = N \hat{1} R^{(i,i)}(k) + \sum_{m=1}^{N-1} \hat{\Gamma}_\mathrm{red}^{(m+1)} $. We notice that the $\hat{\Gamma}^{(m+1)}_\mathrm{red} (k)$ operators  count only the   cycles of degree $m$ with ordered indices $i\rightarrow i+1\rightarrow \dots \rightarrow i+m$, which are therefore less than the permutations counted by the corresponding class-sum operator.

We now show that for every $m$ and $k$ the operator in Eq.~\eqref{eq:Gamma_red} commutes with $\hat{\Gamma}^{(2)}$, namely
\begin{equation}
[\hat{\Gamma}^{(2)}, \hGamma^{(m+1)}_\mathrm{red} (k)] = 0,
\label{eq:comm_Gamma2}
\end{equation}
and, consequently, $[\hGamma^{(2)}, \hat{n}_k] = 0$,
 but does not commute with the SU($2$)-symmetric Hamiltonian, namely
\begin{equation}
[\hat{H}_\mathrm{SU}, \hGamma^{(m+1)}_\mathrm{red} (k)] \neq  0.
\label{eq:comm_HSU}
\end{equation}

To calculate the commutators in Eqs.~\eqref{eq:comm_Gamma2} and~\eqref{eq:comm_HSU}, we introduce the swap operators~\cite{book:auerbach, art:auerbach1988, art:zhang_wang}
\begin{equation}
\Fop{\mu}{\nu}{i} = \ah^\dagger_{i, \mu} \ah_{i, \nu},
\label{eq:FLie}
\end{equation}
where  $\ah_{i, \nu}$ annihilates a boson with flavor $\nu$ at lattice site $i$ and where we omit the hat symbol for the operators $\Fop{\mu}{\nu}{i}$ to simplify the notation. The operators $\Fop{\mu}{\nu}{i}$ follows the commutation relation
\begin{equation}
    \left[\Fop{\mu}{\nu}{i}, \Fop{\alpha}{\beta}{j}\right]=\delta_{i,j}\left(\Fop{\mu}{\beta}{i} \delta_{\alpha\nu}-\Fop{\alpha}{\nu}{i} \delta_{\mu\beta}\right).
    \label{eq:commF}
\end{equation}
In this notation, the SU($2$) Hamiltonian can be written as
\begin{equation}
    \hat{H}_\mathrm{SU} =  - \sum_{i=1}^{N-1}  \sum_{\mu, \nu} \Fop{\mu}{\nu}{i} \Fop{\nu}{\mu}{i+1}
    \label{eq:HSU_def}
\end{equation}
in units of the nearest-neighbours coupling $J=\alpha/g$ and where we have neglected the constant term. We extend the same notation to represent a cycle permutation as
\begin{equation}
\hat{P}_{(i, i+1, \dots, i+m-1, i+m)} = \sum_{\alpha_0, \alpha_1, \dots, \alpha_m} \Fop{\alpha_0}{\alpha_1}{i}\Fop{\alpha_1}{\alpha_2}{i+1} \cdots \Fop{\alpha_{m-1}}{\alpha_m}{i+m-1}\Fop{\alpha_m}{\alpha_0}{i+m}.
    \label{eq:PfuncF}
\end{equation}

\subsection{Commutator with $\hat{\Gamma}^{(2)}$}\label{sec:G2Gred}
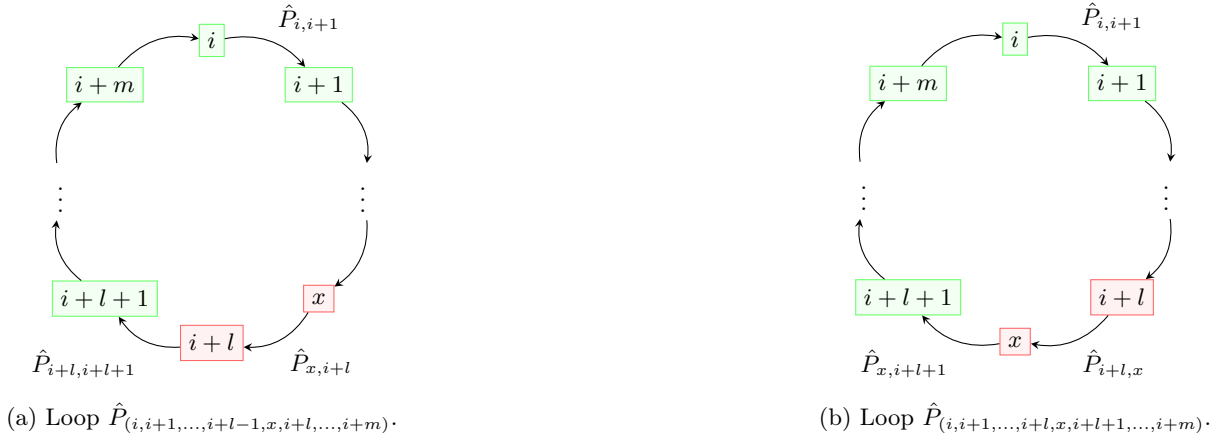
\begin{figure}
\centering
\begin{subfigure}{0.4\textwidth}
\begin{tikzpicture}[node distance=12mm, >=stealth, auto, state/.style={rectangle, draw=green!60, fill=green!5, minimum size=3mm}, state2/.style={rectangle, draw=red!60, fill=red!5, minimum size=3mm}]
\draw (360/8: 2cm) node[state] (i+1) {$i+1$};
\draw (2*360/8: 2cm) node[state] (i) {$i$};
\draw (3*360/8: 2cm) node[state] (i+m) {$i+m$};
\draw (4*360/8: 2cm) node (d1) {$\vdots$};
\draw (5*360/8: 2cm) node[state] (i+l+1) {$i+l+1$};
\draw (6*360/8: 2cm) node[state2] (i+l) {$i+l$};
\draw (7*360/8: 2cm) node[state2] (x) {$x$};
\draw (8*360/8: 2cm) node (d2){$\vdots$};
    \begin{scope}[bend left]
\path[->]   (i+m) edge node {} (i)
            (i) edge node {$\hP_{i, i+1}$} (i+1)
            (i+1) edge node {} (d2)
            (d2) edge node {} (x)
            (x) edge node {$\hP_{x, i+l}$} (i+l)
            (i+l) edge node {$\hP_{i+l, i+l+1}$} (i+l+1)
            (i+l+1) edge node {} (d1)
            (d1) edge node {} (i+m);
    \end{scope}
\end{tikzpicture}
    \caption{Loop $\hP_{(i, i+1, \dots, i+l-1, x, i+l, \dots, i+m)}$.}
    \label{fig:first}
\end{subfigure}
\hfill
\begin{subfigure}{0.4\textwidth}
\begin{tikzpicture}[node distance=12mm, >=stealth, auto, state/.style={rectangle, draw=green!60, fill=green!5, minimum size=3mm}, state2/.style={rectangle, draw=red!60, fill=red!5, minimum size=3mm}]
\draw (360/8: 2cm) node[state] (i+1) {$i+1$};
\draw (2*360/8: 2cm) node[state] (i) {$i$};
\draw (3*360/8: 2cm) node[state] (i+m) {$i+m$};
\draw (4*360/8: 2cm) node (d1) {$\vdots$};
\draw (5*360/8: 2cm) node[state] (i+l+1) {$i+l+1$};
\draw (6*360/8: 2cm) node[state2] (x) {$x$};
\draw (7*360/8: 2cm) node[state2] (i+l) {$i+l$};
\draw (8*360/8: 2cm) node (d2){$\vdots$};
    \begin{scope}[bend left]
\path[->]   (i+m) edge node {} (i)
            (i) edge node {$\hP_{i, i+1}$} (i+1)
            (i+1) edge node {} (d2)
            (d2) edge node {} (i+l)
            (i+l) edge node {$\hP_{i+l, x}$} (x)
            (x) edge node {$\hP_{x, i+l+1}$} (i+l+1)
            (i+l+1) edge node {} (d1)
            (d1) edge node {} (i+m);
    \end{scope}
\end{tikzpicture}
    \caption{Loop $\hP_{(i, i+1, \dots, i+l, x, i+l+1, \dots, i+m)}$.}
    \label{fig:second}
\end{subfigure}
\caption{Schematic representation of the (a) and (b) cycles presented in the commutator $[\hGamma^{(2)}, \hGamma_\mathrm{red}^{(m+1)}]$ (Eq.~\eqref{eq:comm_proof_1}), and labeled according to the terms of Eq.~\eqref{eq:comm_proof_2}.  Every arrow represents a transposition that connects two neighboring sites, represented by the rectangles. The  (c) and (d)  terms of Eq.~\eqref{eq:comm_proof_2} are equivalent to (a) and (b) after the variable change $x \to y$ . }
\label{fig:cycleG2}
\end{figure}

Using Eq.~\eqref{eq:PfuncF} and considering only the loop from $i\to i+m$ (the inverse loop can be calculated using analogous steps), we start the proof of Eq.~\eqref{eq:comm_Gamma2} by writing 
\begin{equation}
    \begin{split}
    [\hGamma^{(2)},  \hGamma^{(m+1)}_\mathrm{red} (k)] &=
        \sum_{x=1}^{N-1} \sum_{y=2}^{N} \sum_{i=1}^{N-m} \sum_{\mathrm{flav.}}\Big([\Fop{\mu}{\nu}{x} \Fop{\nu}{\mu}{y},  \Fop{\lambda}{\alpha_1}{i} \Fop{\alpha_1}{\alpha_2}{i+1} \cdots \Fop{\alpha_{m}}{\lambda}{i+m}]\Big) R^{(i, i+m)}(k)     \\   
        &= \sum_{x=1}^{N-1} \sum_{y=2}^{N} \sum_{i=1}^{N-m} \sum_{\mathrm{flav.}}\sum_{l=0}^{m} 
       \Big\{ \left( \prod_{s=0}^{l-1} \Fop{g(s)[1]}{g(s)[2]}{i+s} \right) [\Fop{\mu}{\nu}{x} \Fop{\nu}{\mu}{y},  \Fop{\alpha_{l}}{\alpha_{l+1}}{i+l}] \\
        &\times 
        \left( \prod_{t=l+1}^{m} \Fop{h(t)[1]}{h(t)[2]}{i+t} \right)\Big\} R^{(i, i+m)}(k),
    \end{split}
    \label{eq:comm_proof_1}
\end{equation}
where we have used the linearity of the commutator and  where $g(s) = \{(\lambda, \alpha_1), (\alpha_1, \alpha_2), \cdots, (\alpha_{l-1}, \alpha_l)) \}$ and $h(t) = \{(\alpha_{l+1}, \alpha_{l+2}), (\alpha_{l+2}, \alpha_{l+3}), \cdots, (\alpha_{m}, \lambda)) \}$ are bi-dimensional arrays containing the two pairs of indices for every $s$ or $t$, and the sum over \qte{flav.} means that all flavors (labeled by the Greek letters) are summed. We now can expand the commutator in the middle of  Eq.~\eqref{eq:comm_proof_1} for the $l$th index in  the following 4 terms
\begin{equation}
\begin{split}
      [\Fop{\mu}{\nu}{x} \Fop{\nu}{\mu}{y},  \Fop{\alpha_{l}}{\alpha_{l+1}}{i+l}]&=
      \delta_{y, i+l}\Big(
    \underbrace{\Fop{\alpha_{l}}{\nu}{x} \Fop{\nu}{\alpha_{l+1}}{i+l} \delta_{\mu, \alpha_l}}_{\mathrm{(a)}} 
        -\underbrace{\Fop{\alpha_l}{\mu}{i+l} \Fop{\mu}{\alpha_{l+1}}{x}  \delta_{\nu, \alpha_{l+1}}}_{\mathrm{(b)}} \Big)
        \\&+\delta_{x, i+l} \Big( \underbrace{\Fop{\alpha_l}{\mu}{y} \Fop{\mu}{\alpha_{l+1}}{i+l} \delta_{\nu, \alpha_l}}_{\mathrm{(c)}} 
        - \underbrace{\Fop{\alpha_l}{\nu}{i+l} \Fop{\nu}{\alpha_{l+1}}{y}  \delta_{\mu, \alpha_{l+1}}}_{\mathrm{(d)}}\Big),
        \end{split}
    \label{eq:comm_proof_2}
\end{equation}

where the (c) and (d) terms are identical to (a) and (b) after the change of variable $y \to x$. The (a) and (b) terms are schematically represented in Fig.~\ref{fig:cycleG2}.   For the cycle, the $l$th (a) term is equal in modulus and opposite in sign to the $(l-1)$th (b)  term and the same happens for the $l$th (b) term and the $(l+1)$th (a) terms. Therefore, they all cancel each other out in neighboring pairs except for the (a)  term with $l=0$ and the (b) term with $l=m$, which they cancel each other out due to the property of the cycle, indeed they correspond to the same cycle, namely $\hat{P}_{(x, i, i+1, \dots, i+m)} = \hat{P}_{(i, i+1, \dots, i+m, x)}$. 

We can repeat the same procedure for the terms dependent on  $y$ and the inverse cycles. Therefore, the commutator in Eq.~\eqref{eq:comm_Gamma2} goes to zero for every $m$  and without adding any constraints on the orbital part $R^{(i, i+m)}(k)$.

\subsection{Commutator with $\hat{H}_\mathrm{SU}$}\label{sec:HGred}
We now focus on Eq.~\eqref{eq:comm_HSU}, where  $\hat{H}_\mathrm{SU}$ is defined in Eq.~\eqref{eq:HSU_def}. We notice that the case $m=1$ is trivial and corresponds to the commutator of the Hamiltonian with itself. In order to compute the commutator, we follow the same procedure used above and we write
\begin{equation}
    \begin{split}
    [\hat{H}_\mathrm{SU}, \hat{\Gamma}^{(m+1)}_\mathrm{red} (k)] &= -
        \sum_{x=1}^{N-1} \sum_{i=1}^{N-m} \sum_{\mathrm{flav.}}\Big([\Fop{\mu}{\nu}{x} \Fop{\nu}{\mu}{x+1},  \Fop{\lambda}{\alpha_1}{i} \Fop{\alpha_1}{\alpha_2}{i+1} \cdots \Fop{\alpha_{m}}{\lambda}{i+m}]  \Big) R^{(i, i+m)}(k)     \\  
        &= -\sum_{x=1}^{N-1} \sum_{i=1}^{N-m} \sum_{\mathrm{flav.}}\sum_{l=0}^{m} 
       \Big\{ \left(\prod_{s=0}^{l-1} \Fop{g(s)[1]}{g(s)[2]}{i+s} \right) [\Fop{\mu}{\nu}{x} \Fop{\nu}{\mu}{x+1},  \Fop{\alpha_{l}}{\alpha_{l+1}}{i+l}] )\\
        &\times 
        \left( \prod_{t=l+1}^{m} \Fop{h(t)[1]}{h(t)[2]}{i+t} \right) \Big\} R^{(i, i+m)}(k),
    \end{split}
    \label{eq:Hcomm_proof_1}
\end{equation}
where the Greek indices are defined as in Eq.~\eqref{eq:comm_proof_1} and the $l$th commutator in the middle gives
\begin{equation}
    \begin{split}
      [\Fop{\mu}{\nu}{x} \Fop{\nu}{\mu}{x+1},  \Fop{\alpha_{l}}{\alpha_{l+1}}{i+l}]=
        \delta_{x, i+l-1}\Big(\underbrace{\Fop{\alpha_{l}}{\nu}{i+l-1} \Fop{\nu}{\alpha_{l+1}}{i+l}  \delta_{\mu, \alpha_l}}_{\mathrm{(a)}} 
        -\underbrace{\Fop{\alpha_l}{\mu}{i+l} \Fop{\mu}{\alpha_{l+1}}{i+l-1}  \delta_{\nu, \alpha_{l+1}}}_{\mathrm{(b)}} \Big)
        \\ + \delta_{x, i+l} \Big(\underbrace{\Fop{\alpha_l}{\mu}{i+l+1} \Fop{\mu}{\alpha_{l+1}}{i+l}\delta_{\nu, \alpha_l}}_{\mathrm{(c)}} 
        - \underbrace{\Fop{\alpha_l}{\nu}{i+l} \Fop{\nu}{\alpha_{l+1}}{i+l+1} \delta_{\mu, \alpha_{l+1}} }_{\mathrm{(d)}} \Big),
    \end{split}
    \label{eq:Hcomm_proof_2}
\end{equation}
whose terms are represented in Fig.~\ref{fig:cycles_H}. It is evident from the picture that the cycles are more complicated than the ones in Fig.~\ref{fig:cycleG2}.
We first consider the middle terms of the sum over $l$, namely $l=1, \dots, m-1$ and show that the (a)-(b) terms  go to zero at given $l$. We will consider the remaining terms, which we call  {\it border terms}, at the end of the section.

\begin{figure}
\centering
\begin{subfigure}{0.4\textwidth}
\begin{tikzpicture}[node distance=12mm, >=stealth, auto, state/.style={rectangle, draw=green!60, fill=green!5, minimum size=3mm}, state2/.style={rectangle, draw=red!60, fill=red!5, minimum size=3mm}]
\draw (360/8: 2cm) node[state] (i+1) {$i+1$};
\draw (2*360/8: 2cm) node[state] (i) {$i$};
\draw (3*360/8: 2cm) node[state] (i+m) {$i+m$};
\draw (4*360/8: 2cm) node (d1) {$\vdots$};
\draw (5*360/8: 2cm) node[state] (i+l+1) {$i+l+1$};
\draw (6*360/8: 2cm) node[state2] (i+l) {$i+l$};
\draw (7*360/8: 2cm) node[state2] (i+l-1) {$i+l-1$};
\draw (8*360/8: 2cm) node (d2){$\vdots$};
    \begin{scope}[bend left]
\path[->]   (i+m) edge node {} (i)
            (i) edge node {$\hP_{i, i+1}$} (i+1)
            (i+1) edge node {} (d2)
            (d2) edge node {} (i+l-1)
            (i+l-1) edge node {$\hP_{i+l-1, i+l}$} (i+l)
            (i+l) edge node {$\hP_{i+l, i+l+1}$} (i+l+1)
            (i+l+1) edge node {} (d1)
            (d1) edge node {} (i+m);
    \end{scope}
    \draw[->, rotate=-30]  (i+l-1.east) .. controls +(right:1cm) and  +(up:1cm) ..(i+l-1.east) node [below = 1mm, pos=0.2] {$\hP_{i+l-1, i+l-1}$}; 
\end{tikzpicture}
    \caption{Loop $\hP_{(i, i+1, \dots, i+l-1, i+l, i+l+1, \dots, i+m)}$.}
    \label{fig:a_H}
\end{subfigure}
\hfill
\begin{subfigure}{0.4\textwidth}
\begin{tikzpicture}[node distance=12mm, >=stealth, auto, state/.style={rectangle, draw=green!60, fill=green!5, minimum size=3mm}, state2/.style={rectangle, draw=red!60, fill=red!5, minimum size=3mm}]
\draw (360/8: 2cm) node[state] (i+1) {$i+1$};
\draw (2*360/8: 2cm) node[state] (i) {$i$};
\draw (3*360/8: 2cm) node[state] (i+m) {$i+m$};
\draw (4*360/8: 2cm) node (d1) {$\vdots$};
\draw (5*360/8: 2cm) node[state] (i+l+1) {$i+l+1$};
\draw (6*360/8: 2cm) node[state2] (i+l) {$i+l$};
\draw (7*360/8: 2cm) node[state2] (i+l-1) {$i+l-1$};
\draw (8*360/8: 2cm) node (d2){$\vdots$};
    \begin{scope}[bend left]
\path[->]   (i+m) edge node {} (i)
 			(i) edge node {$\hP_{i, i+1}$} (i+1)
            (i+1) edge node {} (d2)
            (d2) edge node {} (i+l-1)
            (i+l-1) edge node {$\hP_{i+l-1, i+l}$} (i+l)
             (i+l) edge node [above=2mm] {$\hP_{i+l-1, i+l}$} (i+l-1)
            (i+l+1) edge node {} (d1)
            (d1) edge node {} (i+m);
    \end{scope}
        \draw[->]  (i+l-1.north) .. controls +(up:1cm) and  +(up:1cm) ..(i+l+1.north) node [above = 1mm, pos=0.5] {$\hP_{i+l-1, i+l+1}$};         
\end{tikzpicture}
    \caption{Loop $\hP_{(i, i+1, \dots, i+l-1, i+l, i+l-1, i+l+1, \dots, i+m)}$.}
    \label{fig:b_H}
\end{subfigure}
\hfill
\begin{subfigure}{0.4\textwidth}
\begin{tikzpicture}[node distance=12mm, >=stealth, auto, state/.style={rectangle, draw=green!60, fill=green!5, minimum size=3mm}, state2/.style={rectangle, draw=red!60, fill=red!5, minimum size=3mm}]
\draw (360/8: 2cm) node[state] (i+1) {$i+1$};
\draw (2*360/8: 2cm) node[state] (i) {$i$};
\draw (3*360/8: 2cm) node[state] (i+m) {$i+m$};
\draw (4*360/8: 2cm) node (d1) {$\vdots$};
\draw (5*360/8: 2cm) node[state2] (i+l+1) {$i+l+1$};
\draw (6*360/8: 2cm) node[state2] (i+l) {$i+l$};
\draw (7*360/8: 2cm) node[state] (i+l-1) {$i+l-1$};
\draw (8*360/8: 2cm) node (d2){$\vdots$};
    \begin{scope}[bend left]
\path[->]   (i+m) edge node {} (i)
 			(i) edge node {$\hP_{i, i+1}$} (i+1)
            (i+1) edge node {} (d2)
            (d2) edge node {} (i+l-1)
            (i+l) edge node {$\hP_{i+l, i+l+1}$} (i+l+1)
             (i+l+1) edge node[above=2mm] {$\hP_{i+l+1, i+l}$} (i+l)
            (i+l+1) edge node {} (d1)
            (d1) edge node {} (i+m);
    \end{scope}
        \draw[->]  (i+l-1.north) .. controls +(up:1cm) and  +(up:1cm) ..(i+l+1.north) node [above = 1mm, pos=0.5] {$\hP_{i+l-1, i+l+1}$}; 
\end{tikzpicture}
    \caption{Loop $\hP_{(i, i+1, \dots, i+l-1, i+l+1, i+l, i+l+1, \dots, i+m)}$.}
    \label{fig:c_H}
\end{subfigure}
\hfill
\begin{subfigure}{0.4\textwidth}
\begin{tikzpicture}[node distance=12mm, >=stealth, auto, state/.style={rectangle, draw=green!60, fill=green!5, minimum size=3mm}, state2/.style={rectangle, draw=red!60, fill=red!5, minimum size=3mm}]
\draw (360/8: 2cm) node[state] (i+1) {$i+1$};
\draw (2*360/8: 2cm) node[state] (i) {$i$};
\draw (3*360/8: 2cm) node[state] (i+m) {$i+m$};
\draw (4*360/8: 2cm) node (d1) {$\vdots$};
\draw (5*360/8: 2cm) node[state2] (i+l+1) {$i+l+1$};
\draw (6*360/8: 2cm) node[state2] (i+l) {$i+l$};
\draw (7*360/8: 2cm) node[state] (i+l-1) {$i+l-1$};
\draw (8*360/8: 2cm) node (d2){$\vdots$};
    \begin{scope}[bend left]
\path[->]   (i+m) edge node {} (i)
            (i) edge node {$\hP_{i, i+1}$} (i+1)
            (i+1) edge node {} (d2)
            (d2) edge node {} (i+l-1)
            (i+l-1) edge node {$\hP_{i+l-1, i+l}$} (i+l)
            (i+l) edge node {$\hP_{i+l, i+l+1}$} (i+l+1)
            (i+l+1) edge node {} (d1)
            (d1) edge node {} (i+m);
    \end{scope}
    \draw[->, rotate=-30]  (i+l+1.east) .. controls +(right:1cm) and  +(up:1cm) ..(i+l+1.east) node [above = 4mm, pos=0.2] {$\hP_{i+l+1, i+l+1}$}; 
\end{tikzpicture}
    \caption{Loop $\hP_{(i, i+1, \dots, i+l-1, i+l, i+l+1, \dots, i+m)}$.}
    \label{fig:d_H}
\end{subfigure}
    \caption{Schematic representation of the (a)-(d) cycles presented in the commutator $[\hat{H}_\mathrm{SU}, \hGamma_\mathrm{red}^{(m+1)}]$ (Eq.~\eqref{eq:Hcomm_proof_1}), and labeled according to the four terms of Eq.~\eqref{eq:Hcomm_proof_2}. Every arrow represents a transposition that connects two neighboring sites, represented by the rectangles. }
    \label{fig:cycles_H}
\end{figure}
We notice that the $l$th (a) and (d) terms contain an identity operator $\hat{P}_{i+l-1, i+l-1} = \hat{1}$ represented by the rounded link between the same node in Fig.~\ref{fig:cycles_H}. This can be also understood by considering that the operators in Eq.~\eqref{eq:FLie} are constrained by $\sum_\mu \ah_{i, \mu}^\dagger \ah_{i, \mu} = 1$~\cite{art:zhang_wang}, which yield to
\begin{equation}
\sum_{\alpha}\Fop{\gamma}{\alpha}{i+l} \Fop{\alpha}{\nu}{i+l} = 2 \Fop{\gamma}{\nu}{i+l}.
    \label{eq:Fnumb}
\end{equation}
Therefore, the sum (a)+(d) is equal to zero for every $l$. Notice that so far we did not need to make any assumption on the spatial weight $R^{(i, i+m)}(k)$.

The (b) and (c) terms  also contain  an identity operator, indeed
\begin{equation}
\begin{split}
(b) \,\, \hP_{(i, i+1, \dots, i+l-1, i+l, i+l-1, i+l+1 , \dots, i+m)} &= \cdots \hP_{i+l-2, i+l-1} ( \hP_{i+l-1, i+l}  \hP_{i+l, i+l-1} ) \hP_{i+l-1, i+l-1} \cdots 
\\&= \hP_{(i, i+1, \dots, i+l-1, i+l+1 , \dots, i+m)},
\end{split}
\label{eq:PII} 
\end{equation}
 and 
\begin{equation}
\begin{split}
(c) \,\, \hP_{(i, i+1, \dots, i+l-1, i+l+1, i+l, i+l+1 , \dots, i+m)} &= \cdots \hP_{i+l-2, i+l-1}  ( \hP_{i+l-1, i+l+1}  \hP_{i+l+1, i+l})  \hP_{i+l, i+l+1} \cdots 
\\&= \hP_{(i, i+1, \dots, i+l-1, i+l+1 , \dots, i+m)},
\end{split}
\label{eq:PIII} 
\end{equation}
where we have used for the terms in parenthesis  that $\hP_{i+l-1, i+l}  \hP_{i+l, i+l-1}  = \hP_{i+l-1, i+l}  \hP^{-1}_{i+l-1, i+l} = \hat{1}$ and, consequently, the two are the same cycle.  Therefore, (b) and (c) terms for $ 0 < l < m $ are equal in modulus and opposite in sign. The same steps can be repeated for the intermediate terms of the inverse cycle. 

Finally, we consider the border terms, namely the terms with $l=0$ and $l=m$. 
From Eq.~\eqref{eq:Hcomm_proof_2}, we find that
\begin{itemize}
    \item the (a) term  with $l=0$ and the (d) term with $l=m$ contain the same cycle if we switch $j \to i+1$ and this will give a term
\begin{equation}
 \sum_{i=1}^{N-m-1} \hP_{i \to i+m+1}( R^{(i+1, i+1+m)}(k) - R^{(i, i+m)}(k)),
 \label{eq:IIV_0m}
\end{equation}
which is nonzero for different $R^{(i, j)}(k)$. 
    \item The (a) term  with $l=m$ and the (d) term with $l=0$ cancel one another out after using Eq.~\eqref{eq:Fnumb} for the indices $i+1$ and $i+m-1$, respectively. 
    \item The (c) term with $l=0$ is related to the cycle  $\hP_{i+1, i} \hP_{i, i+1} \hP_{i+1, i+2} \dots \hP_{i+m-1, i+m} = \hP_{i+1 \to i+m}$ where we have used $\hP_{i+1, i} \hP_{i, i+1} = \hat{1}$ and the (b) term  with $l=m$ contains a
    similar structured cycle $\hP_{i, i+1} \hP_{i+1, i+2} \dots \hP_{i+m-1, i+m} \hP_{i+m, i+m-1}= \hP_{i \to i+m-1}$ where we have used $\hP_{i+m, i+m-1} \hP_{i+m-1, i+m} = \hat{1}$.  Therefore, if we switch $j \to i-1$ we can write them in terms of the same cycle $\hP_{i+1 \to  i+m}$.  By also taking into account  additional terms from the extremes of the sum, we obtain the contribution
 \begin{equation}
 \sum_{i=1}^{N-m-1} \hP_{i+1 \to i+m}(R^{(i, i+m)}(k) - R^{(i+1, i+m+1)}(k)) + \hP_{N-m+1 \to N}R^{(N-m, N)}(k) - \hP_{1 \to m}R^{(1, 1+m)}(k),
 \label{eq:III_II_0m}
\end{equation}
which is also nonzero in general.

\item   Finally,  the remaining terms: (b) with $l=0$ and (c) with $l=m$ give another nonzero  contribution, namely
 \begin{equation}
 \sum_{i=1}^{N-m}(\hP_{(i, i+1, \dots, i+m-1, i+m+1,  i+m)  -\hP_{(i, i-1, i+1, i+2, \dots, i+m)})}  R^{(i, i+m)}(k).
 \label{eq:II_III_0m}
\end{equation}  
\end{itemize}
The part related to the inverse cycle of Eq.~\eqref{eq:Hcomm_proof_1} can be derived following the same steps. Finally, the result of the commutator, including cycle and inverse cycle (see Eq.~\eqref{eq:Gamma_red}), gives
\begin{equation}
\begin{split}
[\hat{H}_\mathrm{SU}, \hat{\Gamma}_\mathrm{red}^{(m+1)}] &= \sum_{i=1}^{N-m-1} \left(\hP_{i+m+1\to i} + \hP_{i+1 \to i+m} -\hP_{i\to i+m+1} - \hP_{i+m \to i+1}
   \right)  (R^{(i+1, i+1+m)}(k) - R^{(i, i+m)}(k))\\
&+(\hP_{N \to N-m+1}-\hP_{N-m+1\to N}) R^{(N-m, N)}(k) + (\hP_{1\to m} - \hP_{m \to 1}) R^{(1, m+1)}(k)\\
&- \sum_{i=1}^{N-m}  \Big(\hP_{(i, \dots, i+m-1, i+m+1, i+m)} - \hP_{(i, i-1, i+1, \dots, i+m)} + \hP_{(i+m, \dots, i+1, i-1, i)} \\
&- \hP_{(i+m, i+m+1, i+m-1, \dots, i)}   \Big)  R^{(i, i+m)}(k),
\end{split}
\label{eq:comm_HSU_tot}
\end{equation}
which is nonzero in general. 

We remark that, in the case of periodic boundary condition, $R^{(i, i+m)}$ depends only on $m$ and  therefore the first-line contribution in Eq.~\ref{eq:comm_HSU_tot} is zero. The second line is zero by shifting $N-m + 1 \to 1$ which it is possible in case of periodic boundary conditions. 
The last remaining contribution in Eq.~\eqref{eq:comm_HSU_tot} can be written as 
\begin{equation}
\begin{split}
&2\sum_{i=1}^{N}  \Big(\hP_{(i, i+1, \dots, i+m-1, i+m+1, i+m)} - \hP^{-1}_{(i, i+1, \dots, i+m-1, i+m+1, i+m)} \Big)  R^{(i, i+m)}(k),
\end{split}
\label{eq:remain_term_comm}
\end{equation}
where we have used parity ($x \to N+1-x$) and the change of variable $j=N+1-m-i$ to collect the terms (Eq.~\eqref{eq:remain_term_comm} is twice Eq.~\eqref{eq:III_II_0m} after this change of variable and the factor $2$ takes into account the cycle and inverse cycle). Therefore, this commutator is also nonzero in case of periodic boundary conditions.

\section{Tan's relation in the spin dynamics} \label{app:CN}

In this Section, we show that the Tan relation that 
 relates the asymptotic behavior of the momentum distribution for large $k$ and the expectation value of the derivative of the Hamiltonian with respect to the inverse of the interaction strength still holds in the nonequilibrium scenario presented in this work. This relation derives from the \qte{adiabatic sweep theorem}, which was first demonstrated by Tan~\cite{art:Tan_2008a, art:Tan_largemom} at thermal equilibrium, but it is not valid in general. Indeed,  deviations from this relation have been especially found for nonequilibrium systems~\cite{art:corson_bohn2016, art:colussi2020cumulant, art:Bouchoule_breakdown, art:Rylands2023}.
 Moreover, as mentioned in the main text, we remark that the finite size of the system can also alter this relation in a box trap with rigid borders~\cite{art:aupetit2023, art:schehr_wigner2021} by adding extra terms in the $k^4$ tail which are dependent on the size of the system. Those additional terms, called $\mathcal{C}_b$ in the the main text, do depend on $L$ and $N$ and eventually oscillate in momentum space, but do not contribute to the symmetry oscillations studied in this work and therefore we will neglect them in this section. A study of these oscillations for the spin-mixing dynamics will be the subject of an upcoming publication. 
 
Our goal is to demonstrate that, in the large $k$ limit and in the presence of smooth confinement, the total momentum distribution multiplied by $k^4$ behaves as
\begin{equation}
  \lim_{k\to \infty} k^4 n(k, t)  = \mathcal{C}_T(t)= - \frac{m^2}{\pi \hbar^4}
  \left(\braket{ \Psi(t) |\frac{\partial \hat{H}}{\partial 1/g_{\up\down}} | \Psi(t)}+\sum_{\sigma=\up, \down} \braket{ \Psi(t) |\frac{\partial \hat{H}}{\partial 1/g_{\sigma\sigma}} | \Psi(t)}\right) ,
\label{eq:contact_relation}
\end{equation}
 where   $\ket{\Psi(t)}$ is the many-body wave function  (Eq.(2) of the main text in Dirac notation). 

First, we look at the left side of Eq.~\eqref{eq:contact_relation} and we rewrite the momentum distribution  as
\begin{equation}
\begin{split}
n(k, t) &= \sum_{i, j=1}^N c^{(i, j)} (t) R^{(i, j)}(k) 
 \\&=\frac{N!}{2\pi}\sum_{i, j} c^{(i, j)} (t) \int_{I_{ij}}\left(\prod_{n\neq i}^N dx_n \right) \left|
\int dx e^{-i k x} \Psi_A(x_1 \cdots, x_{i-1}, x, x_{i+1}, \cdots x_N)\right|^2, 
\end{split}
\label{eq:nksigma}
\end{equation}
where $I_{ij} = \{ x_1 < \cdots < x_{i-1} < x < x_{i+1} < \cdots < x_{j} < x' < x_{j+1} < \cdots < x_N \}$ is the integration range  defined in the main text and the spin coefficients $c^{(i, j)}(t)$ are given by
\begin{equation}
    c^{(i,j)} (t)= \bra{\chi(t)} \hat{P}_{(i, \dots, j)}   \ket{\chi(t)}.
    \label{eq:cij}
\end{equation}
 In the large-$k$ limit, the Fourier transform is dominated by the contribution of the short-distance divergence coming from the contact condition~\cite{art:olshanii_PRL2003, art:general_fermions_werner, art:general_bosons_werner}. In this limit, the strongly interacting many-body wave function can be written in the form  
 \begin{equation}
\Psi_A(\vec{X}) \stackrel{|x_i - x_j|\to 0}{\approx} |x_i-x_j| \mathcal{A}\left(\frac{x_i+x_j}{2}, x_{n\neq i, j}\right),
\label{eq:PsiA_exp}
\end{equation}
where  $|x_i - x_j|$ is the two-body wave function satisfying the two-body Schr\"odinger equation in the Tonks limit and $\mathcal{A}$ is a regular function of the center of mass of $x_i$ and $x_j$ and the rest of the coordinates. Importantly, we notice that here we use the Tonks-Girardeau basis instead of the Slater determinant to include the singular behavior at short distances directly in $\Psi_A$.
 One can then use Eq.~\eqref{eq:PsiA_exp} to evaluate the Fourier transform in Eq.~\eqref{eq:nksigma}:
\begin{equation}
\begin{split}
 \int dx e^{-i k x} \Psi_A(x_1 \cdots x \cdots x_N) &\stackrel{|x - x_l|\to 0}{\sim}    \int dx \sum_{l\neq i}  e^{-i k x} \mathcal{A} (x, x_{n\neq i, l}) |x-x_l| \\
 &= -\frac{2}{k^2}  \sum_{l\neq i}^N \mathcal{A} (x, x_{n\neq i, l})   e^{-i k x_l},
\label{eq:int_nk}
\end{split}
\end{equation} 
where we have used that $\lim_{k\to \infty}\int dx e^{-ik x} f(x) = 2 \cos(\pi(\alpha +1)/2) \Gamma(\alpha + 1) \mathrm{exp}(-ik x_0) F(x_0)/|k|^{\alpha+1} + o (1/|k|^{\alpha +2})$ if $f(x)$ has a singularity of the type $|x-x_0|^\alpha F(x)$  at $x_0$ and $F(x)$ is analytic and $\alpha> -1, \alpha \neq 0, 2, 4, \cdots$~\cite{art:olshanii_PRL2003, art:Patu2017}. Notice that the assumption that at large $k$ only short-distance correlations count is not always guaranteed out-of-equilibrium, where the correlations can spread in time, see Ref.~\cite{art:corson_bohn2016} for deep quenches. 

By inserting Eq.~(\ref{eq:int_nk}) into Eq.~(\ref{eq:nksigma}), we find that
\begin{equation}
\begin{split}
\lim_{k\to \infty} k^4 n(k, t) &= \frac{N!}{2\pi}\sum_{i,j} c^{(i, j)} (t) \int_{I_{ij}}\left(\prod_{n\neq i}^N dx_n \right) 4  \sum_{l\neq i}\sum_{m \neq i}
\mathcal{A}^\ast (x, x_{n\neq i, l}) \mathcal{A} (x', x_{n\neq i, m})   e^{-i k (x_m-x_l)} \\
&\stackrel{|x-x'|\to 0}{\sim}\frac{2 N!}{\pi} \sum_{i=1}^{N-1} \left(c^{(i, i+1)}(t) + c^{(i, i)}(t)\right) \sum_{l \neq i} \int_{K_{i}}\left(\prod_{n\neq i}^N dx_n \right)
 |\mathcal{A} (x, x_{n\neq i, l})|^2,
\end{split}
\label{eq:nksigma_fin}
\end{equation}
where the off diagonal terms ($l\neq m$), which are multiplied by a product of regular functions, $\mathcal{A}^\ast \mathcal{A}$, are neglected because they vanish faster than $1/k^4$ due to Riemann -- Lebesgue lemma~\cite{art:Patu2017}, $K_i = \{x_1 < \dots < x_{i-1} < x \leq x' < x_{i+1} < \dots < x_N\}$ such that the only $j$ that survives in the sum are $j=i$ ($x=x'$) and $j=i+1$ ($x \lesssim  x'$). 
Therefore, as mentioned in the main text, Eq.~\eqref{eq:nksigma_fin} only depends on short distance correlations, which means that, at large $k$,  only the terms with $m=0$ and $1$ contributes in Eq.~\eqref{eq:Gamma_red} and $R^{(i, i+m)}(k)$ does not depends on $i$, and therefore $\hat{n}_k$ depends only on $\hat{\Gamma}_\mathrm{red}^{(1)}+\hat{\Gamma}_\mathrm{red}^{(2)}$, whose second term is proportional to the Hamiltonian itself (and therefore, they can commute).

We now demonstrate the right side of Eq.~\eqref{eq:contact_relation}, namely Eq.~\eqref{eq:nksigma_fin} is equal to the Tan contact. To do so, we consider the spin Hamiltonian in Eq.~(3) of the main text, where the spatial part of the wave function is already integrated out. Therefore, one can write the right side of Eq.~\eqref{eq:contact_relation} as
\begin{equation}
\begin{split}
 - \frac{m^2}{\pi \hbar^4}
  \left(\braket{ \chi(t) |\frac{\partial \hat{H}_s}{\partial 1/g_{\up\down}} | \chi(t)}+\sum_{\sigma=\up, \down} \braket{ \chi(t) |\frac{\partial \hat{H}_s}{\partial 1/g_{\sigma\sigma}} | \chi(t)}\right) &=
  \frac{m^2}{\pi \hbar^4} \sum_{i=1}^{N-1} 2\alpha_i  \bra{\chi(t)} \Big(\bf{S}_{i}\cdot\bf{S} _{i+1}  +\frac{3}{4} \Big)\ket{\chi(t)}\\
  &= \frac{m^2}{\pi \hbar^4} \sum_{i=1}^{N-1} \alpha_i  \braket{\chi(t)| \left(\hP_{i, i+1} + 1 \right)| \chi(t)}\\
   &= \frac{m^2}{\pi \hbar^4} \sum_{i=1}^{N-1} \alpha_i \left( c^{(i, i+1)}(t) + c^{(i, i)}(t)\right),
   \end{split}
   \label{eq:adiab_time}
\end{equation}
where, in the second line, we have used that the definition of the swapping operator in terms of spin operators, namely  $\hP_{i, i+1} = 2 {\bf S}_i \cdot {\bf S}_{i+1} + 1/2$~\cite{art:zhang_wang} and, in the third line, we have introduced the spin coefficients (see Eq.~\eqref{eq:cij}).
By recalling the derivation of Ref.~\cite{art:deuretzbacher2014}, the spatial part, $\alpha_i$ (Eq.~(4) of the main text), can be rewritten as
 \begin{equation}
 \begin{split}
 \alpha_i  
 &= \lim_{g_{\sigma\sigma'} \to \infty} g_{\sigma\sigma'}^2 \frac{N!}{2}\sum_{l \neq i} \int d \vec{X} \delta(x_i - x_l) \theta_\mathrm{Id}(\vec{X}) \theta_{P}(\vec{X}) \Psi_A^\ast(\vec{X})\Psi_A^\ast(\vec{X})\\
 &=\lim_{g_{\sigma\sigma'} \to \infty} g_{\sigma\sigma'}^2 \frac{N!}{2} \sum_{l \neq i} \int_{K_i} \prod_{n\neq i} dx_n  \Psi_A^\ast (x_1, \dots, x_i=x_l, \dots x_i=x_l, \dots, x_N) \Psi_A (x_1, \dots, x_i=x_l, \dots x_i=x_l, \dots, x_N),
 \end{split}
 \label{eq:alphai_der}
 \end{equation}
where $K_i$ is the same integration interval of Eq.~\eqref{eq:nksigma_fin} except for $x \to x_i$ and $x' \to x_j$. Then, if one considers the cusp condition~\cite{art:minguzzi2022strongly}, namely
\begin{equation}
\begin{split}
\frac{mg_{\sigma\sigma'}}{\hbar^2} \Psi_A|_{x_i=x_l} &= \frac{\partial \Psi_A}{\partial (x_{i}-x_{l})}\Big|^{0+}_{0-}\stackrel{|x_i-x_l|\to 0}{\approx}  2 \mathcal{A}\left(x_i, x_{n\neq i, l}\right),
\end{split}
\label{eq:cusp_rel}
\end{equation}
where the function $\mathcal{A}$ derives from the short-range expansion in Eq.~\eqref{eq:PsiA_exp}, Eq.~\eqref{eq:alphai_der} becomes
 \begin{equation}
 \begin{split}
 \alpha_i  = \frac{2 N!\hbar^4}{m^2} \sum_{l \neq i} \int_{K_i} \prod_{n\neq i} dx_n \Big|\mathcal{A}\left(x_i, x_{n\neq i, l}\right)\Big|^2.
 \end{split}
 \label{eq:alphai_fin}
 \end{equation}
 Finally, by inserting Eq.~\eqref{eq:alphai_fin} in Eq.~\eqref{eq:adiab_time}, we find
 \begin{equation}
 \mathcal{C}_T(t) = \frac{2 N!}{\pi} \sum_{i=1}^{N-1}\left( c^{(i, i+1)}(t) + c^{(i, i)}(t)\right) \sum_{l \neq i} \int_{K_i} \prod_{n\neq i} dx_n \Big|\mathcal{A}\left(x_i, x_{n\neq i, l}\right)\Big|^2,
     \label{eq:Ct_fin}
 \end{equation}
which is equivalent to Eq.~\eqref{eq:nksigma_fin}.

\section{Analytical expressions for $N=4$}
\label{sec:N=4}

In Fig.(3) of the main text we compare the symmetry oscillations of the expectation value of $\hGamma^{(2)}$ with the one of the momentum distribution at fixed $k$ for $N=6$ in a box potential. The two observables do not oscillate exactly with the same frequencies  due to the fact that the eigenvalues of $\hGamma^{(2)}$, $\gamma_\ell$, are degenerate contrary to the eigenvalues of $\hat{n}_k$, $n_\ell(k)$. For the sake of illustration, we provide here the analytical expressions for $\gamma_2(t)$ and $\mathcal{C}_T$ for the symmetry-breaking case of $4$ particles. 

 Initially, we write $\hGamma^{(2)}$ (Eq.(7) of the main text) and $\hsu$ (Eq.~\eqref{eq:HSU_def}) in the following basis $\{ \ket{\up\up\down\down}, \ket{\up\down\up\down}, \ket{\down\up\up\down},  \ket{\up\down\down\up}, \ket{\down\up\down\up},\ket{\down\down\up\up}\}$, the so-called \qte{snippet} basis~\cite{art:minguzzi2022strongly}, as following
\begin{equation}
\hat{\Gamma}^{(2)} = 
\begin{pmatrix}
2 & 1 & 1 & 1 & 1 & 0 \\
1 & 2 & 1 & 1 & 0 & 1 \\
1 & 1 & 2 & 0 & 1 & 1 \\
1 & 1 & 0 & 2 & 1 & 1 \\
1 & 0 & 1 & 1 & 2 & 1 \\
0 & 1 & 1 & 1 & 1 & 2 
\end{pmatrix},
\,\,\,\,\,\,\,
\hat{H}_\mathrm{SU} = -
\begin{pmatrix}
5 & 1 & 0 & 0 & 0 & 0 \\
1 & 3 & 1 & 1 & 0 & 0 \\
0 & 1 & 4 & 0 & 1 & 0 \\
0 & 1 & 0 & 4 & 1 & 0 \\
0 & 0 & 1 & 1 & 3 & 1 \\
0 & 0 & 0 & 0 & 1 & 5 
\end{pmatrix}.
\label{eq:G2_HSU_N4}
\end{equation}
The matrices in Eq.~\eqref{eq:G2_HSU_N4} are simultaneously diagonalized using the basis
\begin{equation}
\ket{\gamma_0}=
a_0\begin{pmatrix} 1 \\ 1 \\1 \\1 \\1 \\1 \end{pmatrix}, \,\,\,
\ket{\gamma_1}= 
\begin{pmatrix} -a_1 \\ -a_2 \\0 \\0 \\a_2 \\a_1 \end{pmatrix}, \,\,\,
\ket{\gamma_2}=
\begin{pmatrix} -a_3 \\ a_4 \\a_0 \\a_0 \\a_4 \\-a_3 \end{pmatrix}, \,\,\,
\ket{\gamma_3}=
a_5
\begin{pmatrix} 0 \\ 0 \\1 \\-1 \\0 \\0 \end{pmatrix}, \,\,\,
\ket{\gamma_4}=
\begin{pmatrix} a_2 \\ -a_1 \\ 0 \\0 \\a_1 \\-a_2 \end{pmatrix}, \,\,\,
 \ket{\gamma_5}= \begin{pmatrix} a_4 \\ -a_3 \\ a_0 \\a_0 \\-a_3 \\a_4 \end{pmatrix},
\label{eq:gamma_eig}
\end{equation}
with $a_0=1/\sqrt{6}$, $a_1=0.6532$, $a_2=0.2706$, $a_3=0.5577$, $a_4=0.1494$, $a_5=1/\sqrt{2}$. The eigenvectors in Eq.~\eqref{eq:gamma_eig} are ordered  in increasing $\hat{H}_\mathrm{SU}$-eigenvalues, namely
$E_\ell^{SU}=  \{ -6 (\gamma_0=6), -5.41 (\gamma_1=2), -4.73 (\gamma_2=0), -4 (\gamma_3=2), -2.58 (\gamma_4=2),-1.26 (\gamma_5=0) \} $ and where $\gamma_\ell$ in parenthesis are the corresponding $\hat{\Gamma}^{(2)}$-eigenvalues.

We then decompose the spin state in the energy basis, namely $\ket{\chi(t)} =\sum_n s_n e^{-iE_n t} \ket{\chi_n}$ with $s_n=\braket{\chi_n|\chi(0)}$ real coefficients and $\ket{\chi_n}$ eigenvectors of  $\hsb$, which reads, in the snippet basis, as
\begin{equation}
\hsb = -
\begin{pmatrix}
1 & 1 & 0 & 0 & 0 & 0 \\
1 & 3 & 1 & 1 & 0 & 0 \\
0 & 1 & 2 & 0 & 1 & 0 \\
0 & 1 & 0 & 2 & 1 & 0 \\
0 & 0 & 1 & 1 & 3 & 1 \\
0 & 0 & 0 & 0 & 1 & 1 
\end{pmatrix},
\label{eq:HSB_N4}
\end{equation}
whose eigenvalues are $E_n =  \{ -4.73, -3.41, -2, -1.26, -0.58, 0\}$.  The eigenvectors $\ket{\chi_n}$ can be expressed in terms of the basis in Eq.~\eqref{eq:gamma_eig} as follows
 \begin{equation}
 \begin{split}
\ket{\chi_0}&= -0.911 \ket{\gamma_0}-\frac{1}{3}\ket{\gamma_2} + 0.244\ket{\gamma_5} \\
\ket{\chi_1} &= -\frac{1}{\sqrt{2}} \left(\ket{\gamma_1}+\ket{\gamma_4}\right)\\
\ket{\chi_2} &= -\ket{\gamma_3} \\
\ket{\chi_3} &= -0.244 \ket{\gamma_0}+0.911\ket{\gamma_2}+\frac{1}{3}\ket{\gamma_5}\\
\ket{\chi_4} &= \frac{1}{\sqrt{2}} \left(-\ket{\gamma_1}+\ket{\gamma_4}\right)\\
\ket{\chi_5} &= \frac{1}{3}\ket{\gamma_0} -0.244\ket{\gamma_2} +0.911\ket{\gamma_5} 
\end{split}
\label{eq:chin}
\end{equation}

By projecting into the  $\{\ket{\gamma_\ell}\}$ basis, the evolution of the expectation value of $\hGamma^{(2)}$ can be written as
\begin{equation}
\begin{split}
\gamma^{(2)}(t) &= \sum_{m, n=0}^5 s_n s_m \sum_{\ell=0}^5 \braket{\chi_n | \gamma_\ell} \gamma_\ell \braket{\gamma_\ell | \chi_m} e^{-i(E_n-E_m)t}\\
&= 6 (0.83 s_0^2 + 0.06 s_3^2 +0.11 s_5^2) + 2 (s_1^2+s_2^2+s_4^2) \\
&+ 2.62 s_0 s_3 \cos((E_3-E_0)t)- 3.60 s_0 s_5 \cos((E_5-E_0)t)-0.95 s_3 s_5 \cos((E_5-E_3)t),
\end{split}
\label{eq:gamma2_N4}
\end{equation}
which oscillates in time with three different frequencies $\omega_{30} = E_3-E_0$, $\omega_{50} = E_5-E_0$, and $\omega_{35} = E_3-E_5$. 

We then repeat the same steps for the Tan contact (Eq.~\eqref{eq:contact_relation}), and we obtain 
\begin{equation}
\begin{split}
\mathcal{C}_T(t)  &= \sum_{m, n=0}^5 s_n s_m \sum_{\ell=0}^5 \braket{\chi_n | \gamma_\ell} \lambda_\ell \braket{\gamma_\ell | \chi_m} e^{-i(E_n-E_m)t}\\
&= s_0^2 (0.91^2\lambda_0 +0.33^2 \lambda_2 + 0.24^2 \lambda_5)+s_3^2 (0.24^2\lambda_0 +0.91^2 \lambda_2 + 0.33^2 \lambda_5) \\
&+s_5^2 (0.33^2\lambda_0 +0.24^2 \lambda_2 + 0.91^2 \lambda_5) +(s_1^2 + s_4^2) (0.707^2\lambda_1 +0.707^2 \lambda_4 ) +s_2^2 \lambda_3\\
&+s_0 s_3 \cos((E_3-E_0)t) [ 0.22 \lambda_0 -0.3  \lambda_2 +0.08  \lambda_5] +s_0 s_5 \cos((E_5-E_0)t) [-0.3  \lambda_0+0.07\lambda_2+0.22\lambda_5]\\
&+ s_3 s_5 \cos((E_5-E_3)t) [-0.07\lambda_0-0.22\lambda_2 + 0.3 \lambda_5]+ s_1 s_4 \cos((E_4-E_1)t) [0.707\lambda_1 -0.707\lambda_4 ], 
\end{split}
\label{eq:Ct_4p}
\end{equation}
where $\lambda_\ell = \lim_{k \to \infty} k^4 n_\ell (k) = - m^2/(\pi \hbar^4) \braket{\gamma_\ell | \left(\partial \hat{H}_\mathrm{SB}/\partial g_{\uparrow\downarrow}\right)|\gamma_\ell}$. Eq.~\eqref{eq:Ct_4p}  oscillates with four different frequencies $\omega_{30} = E_3-E_0$, 
$\omega_{50} = E_5-E_0$, $\omega_{35} = E_3-E_5$, and $\omega_{14} = E_4-E_1$. In conclusion, $\mathcal{C}_T(t)$ oscillates with an additional frequency ($\omega_{14}$) with respect to $\gamma^{(2)}(t)$ and this originates from the fact that $\lambda_1 \neq \lambda_4$ (non-degenerate eigenvalues), but $\gamma_1 = \gamma_4 = 2$.

We notice that Eqs.~\eqref{eq:gamma2_N4} and~\eqref{eq:Ct_4p} depend on the initial state through the coefficients $s_n$. For the case considered in the main text, the initial state is a mixed-symmetric state and can be written in terms of the $\hat{\Gamma}^{(2)}$-eigenvectors as
\begin{equation}
\ket{\chi(0)} = 0.40 \ket{\gamma_0}+ 0.65 \ket{\gamma_1}-0.55 \ket{\gamma_2} -0.27 \ket{\gamma_4}+0.15\ket{\gamma_5}.
\label{eq:chi0}
\end{equation}

\begin{figure}[h]
\centering
\includegraphics[scale=0.5]{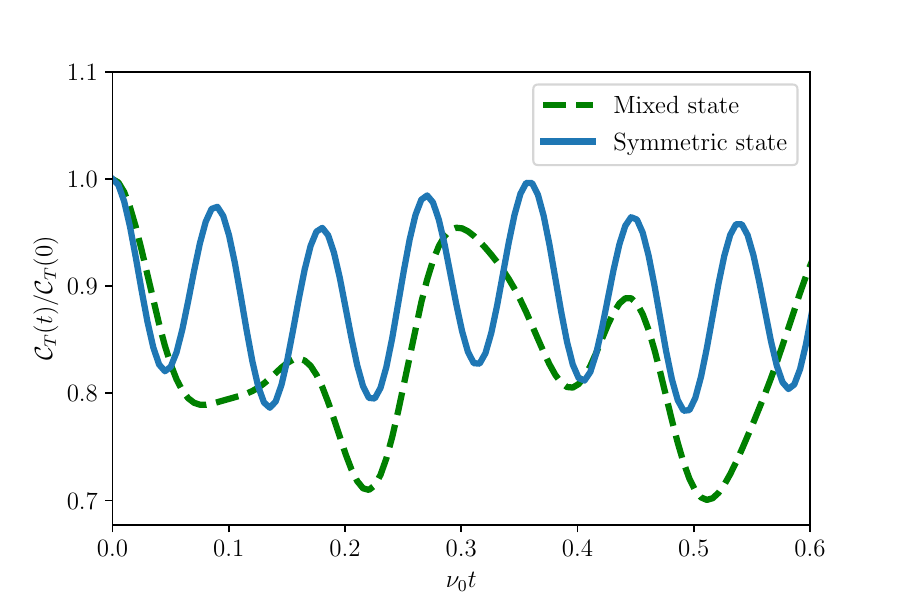}
\caption{Time evolution of the normalized Tan contact for $N=4$ symmetry-breaking bosons for the spin-domain initial state used in the main text and in Eq.~\eqref{eq:chi0} (dashed line) and for the definite-symmetry  initial state $\ket{\chi(0)}=  \ket{\gamma_0}$ (solid line).}
\label{fig:Ct_N4}
\end{figure}

Finally, we consider a different initial state with a specific symmetry, such as $\ket{\chi(0)} = \ket{\gamma_0}$. Whereas for the initial state in Eq.~\eqref{eq:chi0}, all the coefficients $s_n$ are different than $0$, for the latter, only $s_0$, $s_3$, and $s_5$ are nonzero. Both initial states will lead to symmetry oscillations, but the frequencies that matter are not the same, as shown in Fig.~\ref{fig:Ct_N4}.

\bibliographystyle{apsrev4-2}
\bibliography{biblio}